\documentclass[twocolumn,showpacs,preprintnumbers,amsmath,amssymb,prl,unsortedaddress,superscriptaddress,longbibliography,letterpaper]{revtex4-1}
\usepackage{graphicx}% Include figure files
\usepackage[english]{babel}
\usepackage{xfrac}
\begin{document}

%\preprint{APS/123-QED}

\title{Improving the state selectivity of field ionization with quantum control}
%*******************************************************************************
%***Other title options:********************************************************
%*******************************************************************************
%Improving the Rydberg state selectivity of field ionization with quantum control
%	--with a genetic algorithm
%	--using quantum control
%	--using a genetic algorithm
%
%Using a genetic algorithm to improve the Rydberg state selectivity of field ionization
%
%Separating the time-resolved field ionization signals from previously overlapped Rydberg states with a genetic algorithm
%
%Using a genetic algorithm to separate the time-resolved field ionization signals from previously overlapped Rydberg states
%*******************************************************************************
%*******************************************************************************

\author{Vincent C. Gregoric}%
\affiliation{Department of Physics, Bryn Mawr College, Bryn Mawr, PA 19010.}

\author{Jason J. Bennett}
\affiliation{Department of Physics and Astronomy, Ursinus College, Collegeville, PA 19426.}

\author{Bianca R. Gualtieri}
\affiliation{Department of Physics and Astronomy, Ursinus College, Collegeville, PA 19426.}

\author{Ankitha Kannad}%
\affiliation{Department of Physics, Bryn Mawr College, Bryn Mawr, PA 19010.}

\author{Zhimin Cheryl Liu}%
\affiliation{Department of Physics, Bryn Mawr College, Bryn Mawr, PA 19010.}

\author{Zoe A. Rowley}
\affiliation{Department of Physics and Astronomy, Ursinus College, Collegeville, PA 19426.}

\author{Thomas J. Carroll}
\affiliation{Department of Physics and Astronomy, Ursinus College, Collegeville, PA 19426.}

\author{Michael W. Noel}%
\affiliation{Department of Physics, Bryn Mawr College, Bryn Mawr, PA 19010.}

\date{\today}% It is always \today, today,

\begin{abstract}
The electron signals from the field ionization of two closely-spaced Rydberg states of \mbox{rubidium-85} are separated using quantum control. In selective field ionization, the state distribution of a collection of Rydberg atoms is measured by ionizing the atoms with a ramped electric field. Generally, atoms in higher energy states ionize at lower fields, so ionized electrons which are detected earlier in time can be correlated with higher energy Rydberg states. However, the resolution of this technique is limited by the Stark effect. As the electric field is increased, the electron encounters numerous avoided Stark level crossings which split the amplitude among many states, thus broadening the time-resolved ionization signal. Previously, a genetic algorithm has been used to control the signal shape of a single Rydberg state. The present work extends this technique to separate the signals from the $34s$ and $33p$ states of rubidium-85, which are overlapped when using a simple field ramp as in selective field ionization.
\end{abstract}

%\pacs{32.80.Ee, 32.60.+i}% PACS, the Physics and Astronomy

\maketitle
\section{Introduction}
The nearly macroscopic size of highly excited Rydberg atoms has inspired a variety of experiments that explore quantum-classical correspondence.  These include excitation of wavepackets with varying degrees of localization~\cite{ten_wolde_observation_1988,yeazell_classical_1989,naudeau_core_1997,campbell_complete_1999,mestayer_realization_2008,dunning_engineering_2009}, creation of a Schr\"{o}dinger cat-like state~\cite{noel_excitation_1996,chen_dynamics_1997}, and studies in combined electric and magnetic fields, where an equivalent classical system would exhibit chaos~\cite{raithel_quasi-landau_1991,iu_diamagnetic_1991,yeazell_observation_1993,Freund_absorption_2002}.  The large coupling between neighboring Rydberg states allows pairs of atoms to exchange energy in a dipole-dipole interaction~\cite{gallagher_resonant_1992}.  For an ultracold highly-excited sample, many-body effects play an important role in this energy exchange~\cite{anderson_resonant_1998,mourachko_many-body_1998,carroll_many-body_2006}.  An excitation blockade resulting from this strong coupling has also been exploited to entangle atoms and build quantum gates~\cite{jaksch_fast_2000,wilk_entanglement_2010,saffman_quantum_2010,maller_rydberg-blockade_2015,saffman_quantum_2016}.  The large polarizability of Rydberg atoms make them useful for precision measurements of electromagnetic fields~\cite{osterwalder_using_1999,carter_electric-field_2012,facon_sensitive_2016} as well as the quantum state of a nanomechanical oscillator~\cite{sanz-mora_-chip_2017}. 

In many experiments, population is spread among several Rydberg states, either during excitation or by subsequent interactions.  Understanding the dynamics of these Rydberg systems typically requires accurate measurement of the electron's state distribution.  Selective field ionization (SFI) is often used for this purpose \cite{gallagher_field_1977}.  In this technique, an electric field ramp is applied to a sample of Rydberg atoms.  As the field increases, more tightly bound states are ionized.  Therefore, the time-resolved electron signal provides a measure of the distribution of population among Rydberg states, with earlier arrival corresponding to high principal quantum number and later arrival to more tightly bound states.  While this simple picture provides a reasonable qualitative understanding of SFI, the details of the field ionization process complicate the signal, often making neighboring states difficult to resolve.   

A modification to SFI was recently developed in which the electron is directed through the many Stark states it encounters on the way to ionization, thus controlling the shape of the time-resolved signal~\cite{gregoric_quantum_2017}.  This is done by perturbing the electric field ramp with a continuous series of small fluctuations in the electric field. These perturbations manipulate the phase evolution of the Stark states, thus controlling the output amplitudes at each avoided crossing. A genetic algorithm (GA) is used to optimize the perturbation to manipulate the time-resolved signal. 

In this work, we present the results of an experiment in which we use this directed field ionization to separate the signals from two nearby states, the $33p_{3/2,|m_j|=1/2}$ and the $34s$, whose time-resolved signals are almost completely indistinguishable when obtained using traditional SFI. Our choice of states is motivated by the $np_{3/2} + np_{3/2}\rightarrow (n+1)s + ns$ dipole-dipole interaction. Since the $34s$ and $33s$ signals are difficult to resolve from the $33p_{3/2}$, this dipole-dipole energy exchange is challenging to measure.

\begin{figure*}
	\centering
	\includegraphics{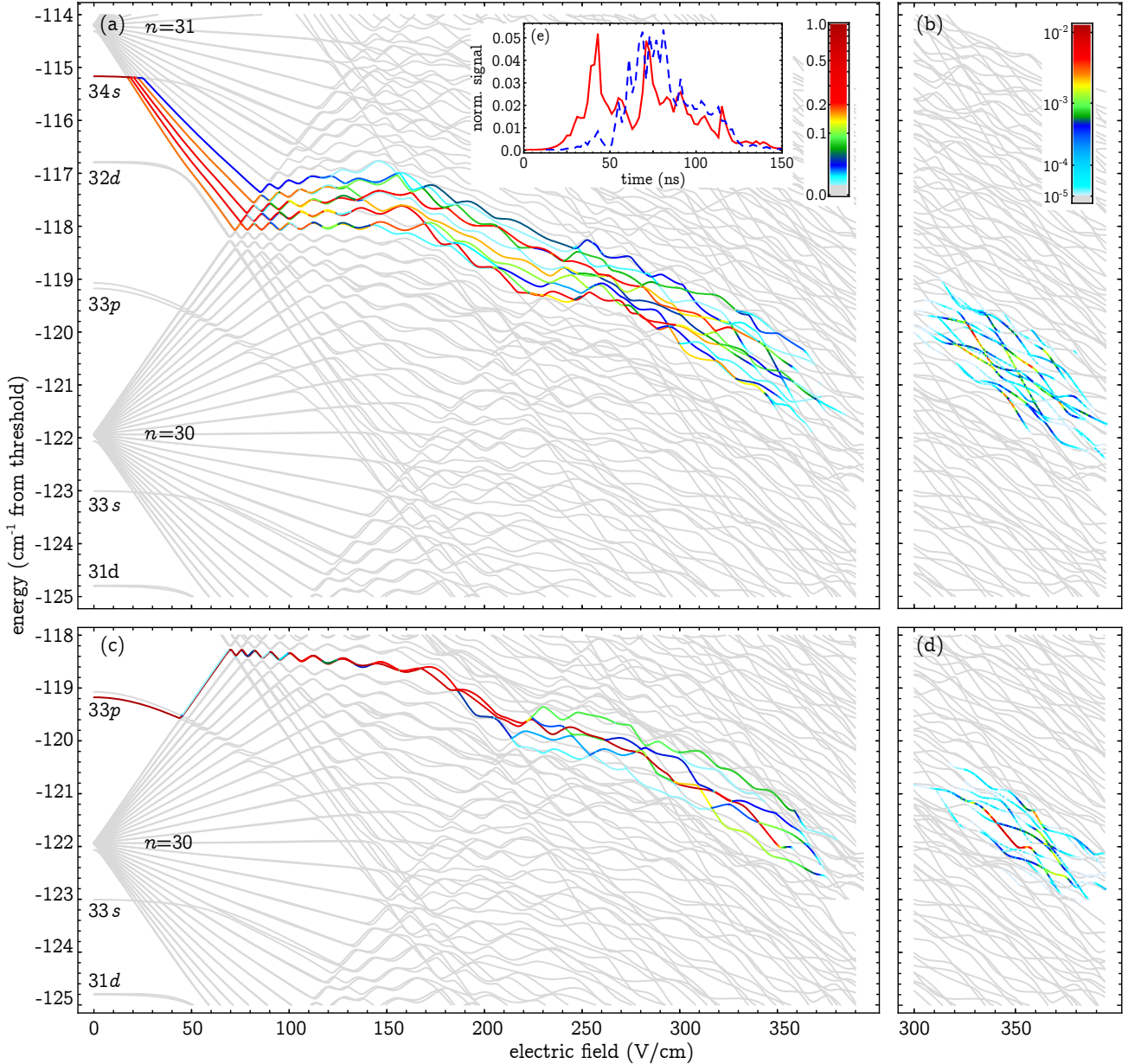}
	\caption{(Color online) Calculated paths to ionization, population ionized from each state, and time-resolved field ionization signals for the unperturbed SFI ramp for $|m_j|=1/2$. The calculation was performed by constructing the time evolution operator using a basis including the Stark states from $n=26$ to $n=36$ with a time resolution of 0.01~ns, following the method previously described in~\cite{feynman_quantum_2015}. The paths to ionization for the $34s$ and $33p$ initial states are shown in (a) and (c), respectively. Each line is colored according to the population remaining in that state using the legend in (a). Note that there is very little overlap among the states populated by the $34s$ path and the $33p$ path. This can be seen by following the $32d$ state. At around 70~V/cm, where the $n=30$ and $n=31$ manifolds collide, the $32d$ state is in between the $34s$ and $33p$ states, neither of which couple significantly to the $32d$ until past 200~V/cm. The population ionized from each state in each 50~ns time interval for the initially populated $34s$ and $33p$ states is shown in (b) and (d) respectively, with each line colored by the legend in (b). Note that in (b) and (d) the color refers to the population \textit{leaving} the state, in contrast to (a) and (c) which show the population remaining in each state. Even though the $34s$ and $33p$ paths spread across a different, and nearly non-overlapping, set of states, they ionize at roughly the same fields. This is seen clearly in the calculated time-resolved field ionization signal shown in (e), where the $34s$ (red, solid) and $33p$ (blue, dashed) signals have a significant overlap of 73.2\%.
	}
	\label{fig:path}
\end{figure*}

Figure~\ref{fig:path}(a) and (c) show the calculated path to ionization using the unperturbed SFI ramp for the $34s$ and $33p$ states, respectively. This field rises to 600~V/cm in 1500~ns, resulting in a slew rate of 0.4~(V/cm)/ns. We will refer to the Stark states by the label of the zero-field state to which they are adiabatically connected. The population in each state is indicated by its color. As the field increases, each state encounters many avoided crossings. This leads to a spreading of population across many states as the ionization threshold is approached, resulting in an ionization signal that is spread out in time.

It is interesting to note that while the population that was initially in the $34s$ and $33p$ states both spread across many states during field ionization, there is not much overlap in the set of states that each populate near threshold. In spite of this, the time-resolved signals for field ionization of the $34s$ and $33p$ states are almost completely overlapped. This can be understood by considering Fig.~\ref{fig:path}(b) and (d), which show how much population has ionized from each state in each 50~ns time interval. Here, the color indicates the population that is ionizing rather than the population remaining. While the states do not overlap, much of the population ionizes over the same range of fields, thus producing the well-overlapped calculated ionization signal in Fig.~\ref{fig:path}(e), which compares favorably to the experimental signals shown in Fig.~\ref{fig:data}(a)~--~(c). In Fig.~\ref{fig:path}(b) and (d) we also see that neighboring states ionize with dramatically different rates.  This is due to the relative orientation of the electron wave function and the electric field, providing the GA with opportunities near threshold to control the timing of ionization.

\section{Experiment}

To experimentally achieve state separation of the $34s$ and $33p$ states, we first confine about a million rubidium-85 atoms in a magneto-optical trap (MOT), which cools the atoms to approximately 200~$\mu$K. Homemade external cavity diode lasers of wavelengths 780~nm, 776~nm, and 1022~nm are used to excite the trapped atoms to the $34s$ state~\cite{fahey_excitation_2011}; for the $33p$ state, a 1270~nm laser is used in place of the 1022~nm laser~\cite{fahey_imaging_2015}. To alternate between exciting the $34s$ and $33p$ states on subsequent shots of the experiment, we tune the 1270~nm and 1022~nm lasers in and out of resonance by adjusting the acoustic frequency of two acousto-optic modulators. %We then field-ionize the rubidium using a linearly increasing electric field with perturbations that are optimized by a GA. The time-resolved ionization signal is used to determine the state distribution of the atoms. 
After excitation, the Rydberg atoms are field ionized and the time-resolved ionization signal is recorded. This experimental cycle is repeated at a 60~Hz rate.

\begin{figure}
	\centering
	\includegraphics{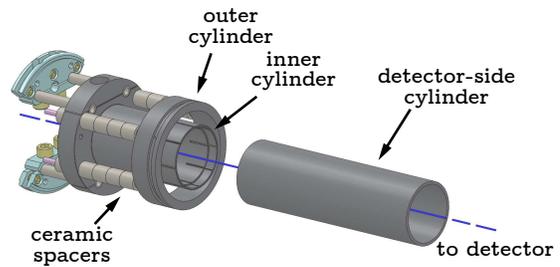}
	\caption{(Color online) Electrode geometry.  The MOT sits on the axis of a set of coaxial cylinders.  Two cylinders, labeled inner and outer, are on one side of the MOT and a third, labeled detector-side, is on the opposite side.  The field ionization ramp is applied to the inner and outer cylinders and the perturbing field is applied to the detector-side cylinder.
	}
	\label{fig:exp}
\end{figure}

The electric field experienced by the atoms is controlled by three coaxial cylindrical electrodes as shown in  Fig.~\ref{fig:exp}. Two concentric cylinders on one end of the trap can be independently biased to control the homogeneity of the electric field. A sufficiently homogeneous field was achieved when equal voltages were applied to these two cylinders. A static DC voltage is applied to these cylinders, producing an electric field of 13~V/cm which allows us to resolve the different $\left|m_j\right|$ sublevels so that we can selectively excite the $33p_{3/2,|m_j|=1/2}$ state. The ionizing field ramp is also applied to these electrodes using a trigger transformer circuit controlled by a MOSFET switch. The perturbing electric field to be optimized by the GA is applied to a third electrode on the opposite end of the trap. The arbitrary waveform generator used to produce this perturbing field has 14-bit resolution, a sample rate of 1~GS/s, and can switch from $+10$~V to $-10$~V (corresponding to electric fields of $\pm3.8$~V/cm) in 3.3~ns. This results in possible slew rates for the combined electric field ranging from $-1.6$~(V/cm)/ns to $+3.0$~(V/cm)/ns. %After ionization, the electron signal is amplified by multi-channel plates before being fed to an oscilloscope. 

The GA starts by generating a population of 120 random electric field perturbations. Each pulse is assigned a fitness score based on how well it achieves the desired outcome; either by moving the arrival times of the $34s$ and $33p$ ionization signals in opposite directions or by reducing the overlap of the two ionization signals. The next generation is populated with the top eight best scoring members of the population along with offspring that are created by mixing the genes, in this case the field values of the perturbation, from the more successful parents. Tournament selection with a tournament size of four is use to select the parents. Each gene is subjected to a 1\% chance of mutating; if a gene mutates, it is reset to a random value. For a fuller description of our algorithm, see~\cite{gregoric_quantum_2017}.

\begin{figure*}
	\centering
	\includegraphics{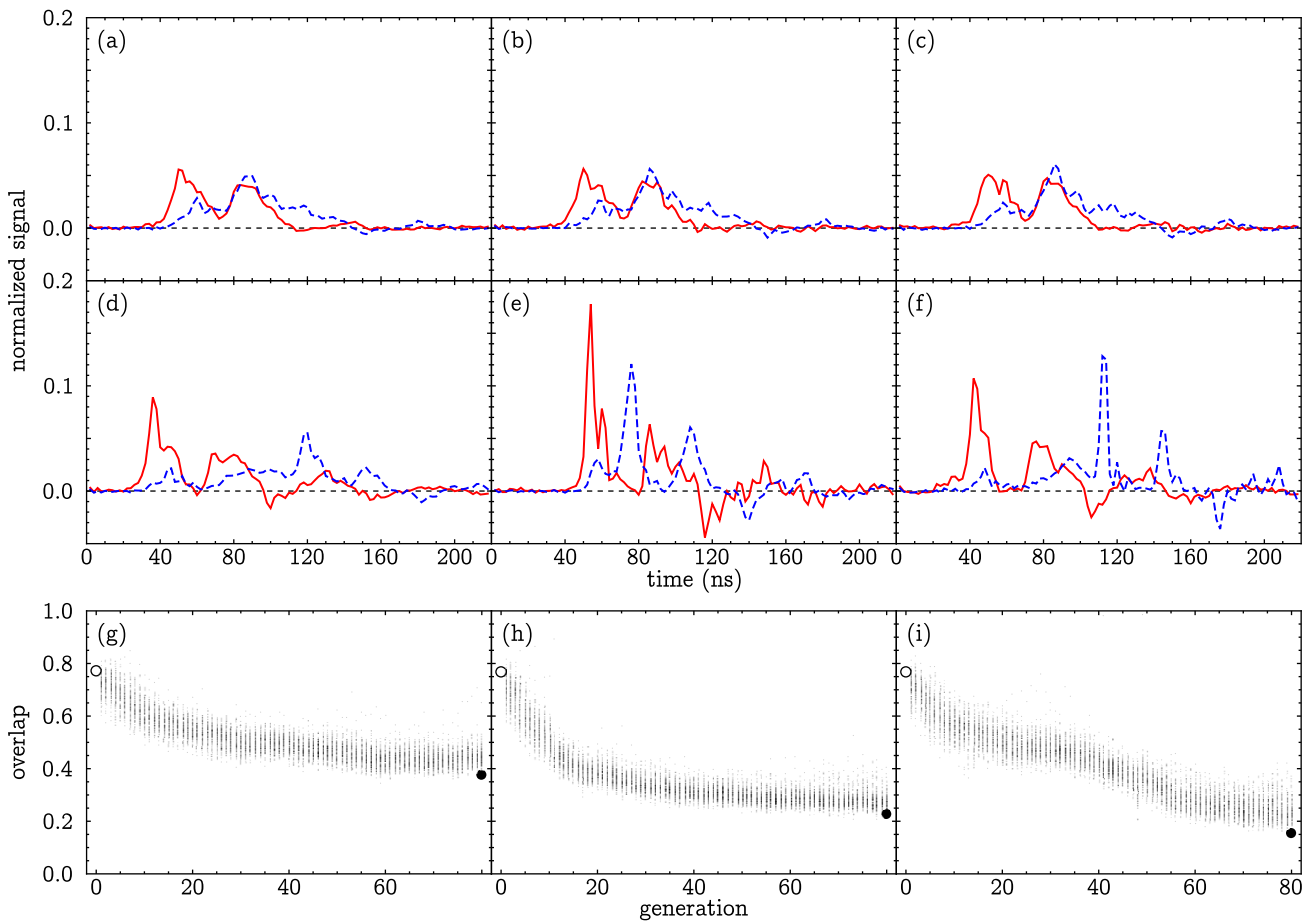}
	\caption{(Color online) GA scans to separate the $34s$ (red, solid) and $33p$ (blue, dashed) states. The unperturbed traces are shown in (a)~--~(c), while the best results from the last generation are shown in (d)~--~(f). In (g)~--~(i), the overlap between the two states is plotted vs.~generation for each member of the GA population; the large, open circles represent the unperturbed overlap, while the large, filled circles show the minimum overlap achieved in the last generation. The left and center columns correspond to GAs using the weighted shift and minimize overlap fitness scores, respectively. For the right column, a weighted shift fitness score was used for the first 40 generations before switching to the minimize overlap fitness score for the remainder of the optimization. 
	}
	\label{fig:data}
\end{figure*}

The performance of a GA is highly dependent on how the fitness score is calculated, since this determines which genetic material is passed down to future generations. We have tested several different methods for calculating fitness scores in the case of two-state separation. One example is a  ``weighted shift'' fitness score, in which the normalized signals from each state are multiplied by a linear weighting function. To shift the $34s$~state to the left or to earlier electron arrival times, this linear weight has a value of one on the left of the total signal gate and a value of zero on the right side of the gate. For shifting the $33p$~state to the right or to later electron arrival times, the weighting function is reflected horizontally (rising linearly from zero on the left side of the gate to one on the right side). The total fitness score is the geometric mean of the weighted signals.

We have used this fitness score to separate the $34s$~and $33p$~states, as shown in the left column of Fig.~\ref{fig:data}. The initial (unperturbed) signals for the~$34s$ state (red, solid) and the~$33p$ state (blue, dashed) are shown in Fig.~\ref{fig:data}(a), while the traces for the best result in the final generation are shown in Fig.~\ref{fig:data}(d). For these traces, the total area under each curve is normalized to one. The overlap between the $34s$~and $33p$~states is plotted in Fig.~\ref{fig:data}(g) as a function of generation for each perturbation tested. The large open and closed circles mark the overlap for the unperturbed case and the best result in the final generation, respectively. Using the weighted shift fitness score, we were able to decrease the overlap between the $34s$~and $33p$~states from~$77.0\%$ to~$37.9\%$.

While the weighted shift fitness score was able to significantly reduce the state overlap, we have made more progress by directly including the overlap into the fitness score calculation. Specifically, we define this fitness score as the difference between one and the overlap integral of the $34s$~and $33p$~signals, so that a smaller overlap corresponds to higher fitness. We calculate the overlap integral of two discrete signal traces by taking the dot-product. Our ``minimize overlap'' fitness score is normalized by dividing the overlap integral by the norms of the signal vectors. The results of running a GA using this fitness score are shown in the center column of Fig.~\ref{fig:data}, following the same conventions used for the left column. Compared to the weighted shift, the minimize overlap fitness score performs significantly better, decreasing the overlap from~$76.8\%$ to~$22.8\%$ over the course of the GA.

One potential issue with the minimize overlap fitness score is that it can result in signals which alternate in time between the~$34s$ and~$33p$ states, such as the interleaved signals in Fig~\ref{fig:data}(e). To avoid this, we have also tested a hybrid ``shift then overlap'' GA, which initially uses the weighted shift fitness score for a fixed number of generations before switching to the minimize overlap fitness score for the remainder of the optimization. This hybrid GA, shown in the right column of Fig.~\ref{fig:data}, outperforms both of the previous datasets, decreasing the overlap from~$76.6\%$ to~$15.4\%$ while avoiding interleaved signals. Note the repeated pattern of decrease and then plateau in the overlap for the hybrid GA in Fig.~\ref{fig:data}(i); this is a result of switching the fitness score from weighted shift to minimize overlap at generation 40.

%We have also attempted to separate the signals from different $\left|m_j\right|$ sublevels of the $32d_{5/2}$ state (data not shown), though this has proven to be more challenging for our GA. One potential explanation for this difficulty is the fact that all of the $32d$ states take a very similar pathway through the avoided crossings, making it hard for the GA to manipulate the signal from two different $32d$ states in different ways. This is in contrast to the ionization pathways for the~$34s$ and~$33p$ states, as discussed above.

\begin{figure}
	\centering
	\includegraphics{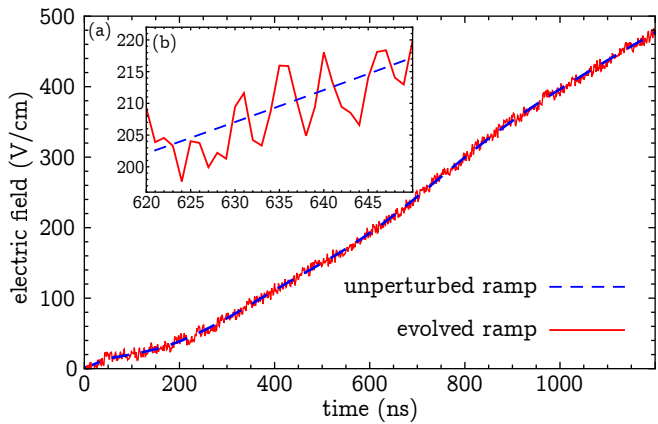}
	\caption{(Color online) (a) The unperturbed electric field ramp (dashed blue) and the optimized ramp (solid red) for the fitness score shown in Fig.~\ref{fig:data}(i). Since the perturbations are quite small on the scale of the whole ramp, a typical region is shown in (b). The perturbations extend through ionization, which is completed by about 400~V/cm for the $34s$ and $33p$ states studied here, and each perturbation is a few V/cm.
	}
	\label{fig:pulse}
\end{figure}

The unperturbed electric field ramp is shown along with one of the optimized ramps in Fig.~\ref{fig:pulse}. While the perturbations are small compared to the size of the ramp, they are sufficient to control the phase along the path to ionization through many avoided crossings. Given the complexity of the Stark map along with the uncertainty in completely characterizing the experimental conditions, it is difficult to correlate the individual fluctuations in the optimized field with particular avoided crossings.

One way to gain some physical insight into the GA optimization process is to probe different regions of the Stark map by altering the duration of the field perturbations between otherwise identical GA scans. We have explored this by taking several datasets using the weighted shift fitness score. In each dataset, we begin the perturbation at an electric field of 9~V/cm which is before both the~$34s$ and~$33p$ states hit the high-$\ell$ manifolds. The perturbation end time is varied between datasets. For GA runs where the perturbation ends before either state hits a manifold, no change is observed in the ionization signals, as expected. If the perturbation is extended past 17~V/cm, corresponding to the point at which the $34s$~state hits the $n=31$ manifold, the GA is then able to shift the $34s$~state signal left. However, the GA is not able to shift the $33p$~state to the right unless the perturbation is extended nearly to the ionization threshold, well past the point at which the $33p$~state hits the $n=30$ manifold at 45~V/cm. This matches the computational analysis of the paths to ionization presented in Fig.~\ref{fig:path}. The~$34s$ state takes a mix of the adiabatic and diabatic pathways through the first few avoided crossings with the $n=31$ manifold, allowing our perturbations to shift this behavior toward either extreme. For the $33p$ state, however, the pathway is strongly adiabatic until $\approx$200~V/cm. As a result, our perturbations are not large enough to significantly shift the $33p$ state's ionization pathway toward the diabatic regime during these early crossings.  

\section{Discussion}
A significant fraction of the success of the GA is due to the details of the ionization process near threshold. The addition of the ionizing ramp potential to the coulomb potential creates a saddle point in the total potential. For electrons of sufficient energy, ionization is classically allowed at the saddle point, while electrons of lower energies can tunnel to ionization~\cite{littman_tunneling_1976}. Each state can be characterized by the spatial distribution of its wavefunction. Higher energy states, in which the electron is on the opposite side of the atom from the saddle point, are harder to ionize and typically referred to as ``blue'' states. Lower energy states, in which the electron is localized to the same side of the atom as the saddle point, are easier to ionize and typically referred to as ``red'' states. 

In nonhydrogenic atoms like rubidium, the red and blue states are coupled by their interaction with the core. Rather than crossing as they do in hydrogen, red and blue states from neighboring $n$ will exhibit avoided crossings~\cite{gallagher_rydberg_1994}. In the region of the Stark map near ionization, coupled states can have dramatically different ionization rates; our calculated ionization rates (using the method of~\cite{damburg_hydrogen_1979}) show that it is easy to find examples of neighboring Stark states with ionization rates differing by more than five orders of magnitude. Adjacent states in Na around $n=13$ have been shown to reach a threshold ionization rate of $10^7$~s$^{-1}$ at fields differing by more than 10~kV/cm~\cite{littman_tunneling_1976}. The ionization rates can also change due to interference between the decay channels at an avoided crossing, an effect studied in the photoionization peaks of Rb~\cite{feneuille_field-induced_1982} and line narrowing in the photoionization spectrum of Na~\cite{liu_interference_1985}. These widely varying ionization rates provide an ideal landscape for the GA, which can choose perturbations that move population into either rapidly- or slowly-ionizing states, depending on whether it is desired to move the ionization signal earlier or later in time. 

\begin{figure*}
	\centering
	\includegraphics{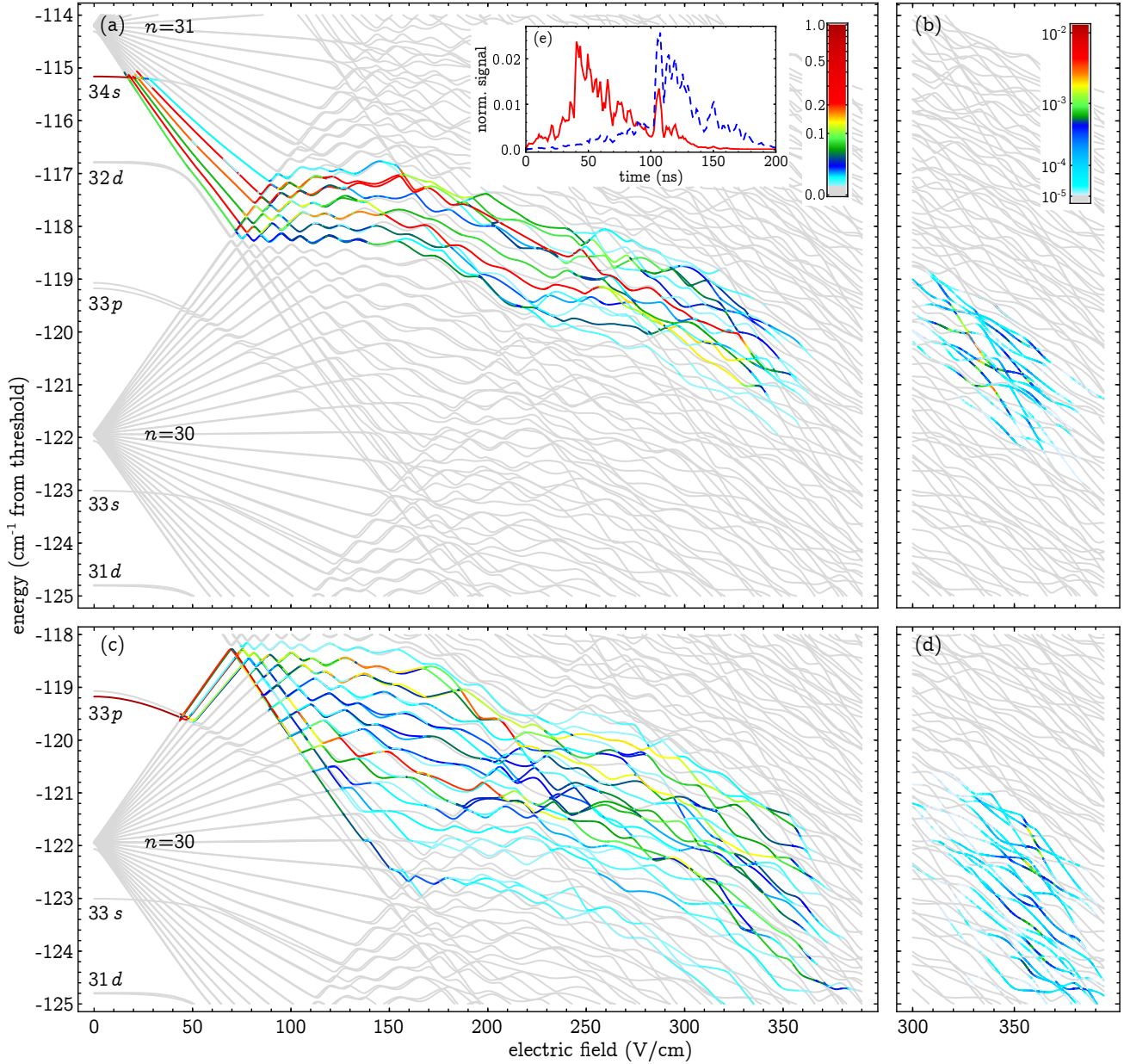}
	\caption{(Color online) Calculated paths to ionization, population ionized from each state, and time-resolved field ionization signals for an evolved ramp for $|m_j| = 1/2$. The ramp was evolved using the same hybrid fitness score as in Fig.~\ref{fig:data}(i), but limited to only 30 total generations due to computational time constraints. The calculation was performed in the same way as for the unevolved paths shown in Fig.~\ref{fig:path}. The evolved paths to ionization for the $34s$ and $33p$ initial states are shown in (a) and (c), respectively. Each line is colored according to the population remaining in that state using the legend in (a), which is the same scale as used in Fig.~\ref{fig:path}(a).  The population ionized from each state in each 50 ns time interval for the initially populated 34s and 33p states is shown in (b) and (d) respectively, with each line colored by the legend in (b), which is the same scale as used in Fig.~\ref{fig:path}(b). Note that in (b) and (d) the color refers to the population leaving the state, in contrast to (a) and (c) which show the population remaining in each state. In comparing these evolved paths to Fig.~\ref{fig:path}, it is clear that the GA has made some effort to push the amplitudes to higher and generally earlier ionizing states for the $34s$ and to lower and generally later ionizing states for the $33p$. However, the local variation in ionization rates among neighboring states is as important as the general trend of higher ionization rates at higher energies. Even though the set of states from which the $33p$ and $34s$ finally ionize do not significantly overlap, there is still an overlap in the field ionization signals as seen in (e). The simulated GA successfully reduces the overlap from 73.2\% in Fig.~\ref{fig:path}(e) to 41.0\%.
	}
	\label{fig:evpath}
\end{figure*} 

We have also simulated the GA by repeating the same calculation as shown in Fig.~\ref{fig:path} in parallel for a population of 48 electric field ramps over 30 generations. The final evolved paths to ionization are shown in Fig.~\ref{fig:evpath} along with the simulated time-resolved signal. The simulated GA reduced the overlap of the $34s$ and $33p$ states from 73.2\% in Fig.~\ref{fig:path}(e) to 41.0\% in Fig.~\ref{fig:evpath}(e). While the states from which the electron amplitude ionizes do not overlap, as shown in Fig.~\ref{fig:evpath}(b) and Fig.~\ref{fig:evpath}(d), there still remains some overlap in the time-resolved signal. This is because the local variation in ionization rates among neighboring states is significant compared to the general trend of higher ionization rates at higher energies. 

Our simulations show that the GA transfers amplitude between slow and fast ionizing states near threshold. We have run simulations to compare perturbations that end much earlier than the ionization region to perturbations that are only present around the ionization region. Similar to the  experimental datasets with varying perturbation length discussed above, the perturbations that are present only around the ionization region perform better. We have determined that about 2/3 of the improvement in fitness score is due to the portion of the perturbations just before and during ionization.

While our model is successful in accounting for many of the observed experimental features and yields information not accessible in the experiment, it cannot be used for more than general guidance for three primary reasons. First, the model is incomplete in the sense that its limited basis includes only bound states. We calculate ionization rates using a semi-empirical formula rather than directly from the couplings to free states. In Feynman, \textit{et al}.\ essentially the same model was unable to correctly account for the phase evolution near ionization~\cite{feynman_quantum_2015}. Second, a significant advantage of the GA is that it automatically takes into account uncharacterized experimental conditions, such as electric and magnetic field inhomogeneity. Both the model and the experiment reveal that small changes in the electric field can have large effects. Since it is not feasible to measure all of the particular experimental conditions, the model cannot calculate a path to ionization that precisely captures the experiment. Finally, the simulated optimization of Fig.~\ref{fig:evpath} takes about 10 days to run on a modern supercomputer. The experiment is far more efficient, completing a similar optimization in only about one hour.

\section{Conclusion}

We have demonstrated the ability of our GA to separate the overlapped ionization signals from the~$34s$ and~$33p$ states of rubidium. By changing the fitness score calculation partway through the GA, we have been able to decrease the state overlap while avoiding interleaved signals. This technique will be useful in experiments requiring differentiation between the~$34s$ and~$33p$ states. Specifically, we plan to use the results of this work to study the dipole-dipole interaction \mbox{$np + np \rightarrow ns + (n+1)s$.}  It should be straightforward to use this technique to separate the signals from other states whose ionizations signals are overlapped when traditional SFI is used.  This optimization technique may be useful for other goals as well.  For example, the production of high-brightness, monochromatic electron beams using field ionized Rydberg atoms may benefit from the addition of an optimized perturbation to the ionizing field \cite{kime_high-flux_2013,mcculloch_field_2017,moufarej_forced_2017}.

This work was supported by the National Science Foundation under Grants No. 1607335 and No. 1607377.
%\bibliographystyle{apsrev4-1}
%\bibliography{directedFieldIonization}% Produces the bibliography via BibTeX.
%\bibliographystyle{abbrv}

\bibliography{GA2State}% Produces the bibliography via BibTeX.

\end{document}